# Double core-hole electron spectroscopy for open-shell molecules: theoretical perspective

Motomichi Tashiro [a,b], Masahiro Ehara [a,b,*], Kiyoshi Ueda [c]

[a] *Institute for Molecular Science, Myodai-ji Nishigo-naka 38, Okazaki 444-8585, Japan*

[b] *Research Center for Computational Science, Myodai-ji Nishigo-naka 38, Okazaki 444-8585, Japan*

[c] *Institute of Multidisciplinary Research for Advanced Materials, Tohoku University, Sendai 980-8577, Japan*

**ABSTRACT**

We have theoretically investigated the double core-hole (DCH) states of the open-shell molecules and examined the possibility of DCH spectroscopy by means of X-ray two-photon photoelectron spectroscopy (XTPPS). Energies of many DCH states were obtained by the CASSCF calculations and the generalized intra- and interatomic relaxation energies were evaluated. We show that XTPPS can extract these quantities by the measurement of single and double core-hole ionization potentials. We discuss the influence of chemical environment on the DCH states with two holes at the same atomic site and at two different atomic sites.

_________________________

* Corresponding author. FAX: +81-564-55-7025

   *E-mail address*: ehara@ims.ac.jp (M. Ehara)

## 1. Introduction

More than two decades ago, Cederbaum et al. investigated double core-hole (DCH) states in molecules, $C_2H_2$, $C_2H_4$, $C_2H_6$ [1,2], $C_6H_6$ [3], $SiH_4$ and $SiF_4$ [4], and provided the fundamental chemical concept of the molecular DCH states. Namely, they analyzed characters of the DCH states with two holes at the same atomic site and those at different atomic sites and found that the DCH states having two holes at different atomic sites are very sensitive to the chemical environment of the two holes. Conventional X-ray photoelectron spectroscopy based on one-photon ionization, however, cannot be accessible to such two-site DCH states because one-photon ionization probability to these two-site states is practically zero, though DCH states with two holes in one atomic site may be accessible via one-photon ionization [5,6].

Experimentally, the two-atomic site DCH states are accessible only by sequential X-ray two-photon absorption. Such X-ray two-photon photoelectron spectroscopy (XTPPS) [7] has not yet been realized so far. Recent developments of short-wavelength free-electron lasers (FEL) [8], however, have been changing the situations very rapidly. At FEL facilities in operations, such as FLASH in Hamburg [9] and SPring-8 Compact SASE Source (SCSS) test accelerator [10,11], multi-photon multiple-ionization processes in molecules have been extensively studied [12-15]. Furthermore, X-ray FEL facility LCLS in Stanford, which has just started operation [16], is expected to generate ~1-fs pulses containing $2.4 \times 10^{11}$ photons at 1 keV [17,18]. Inspired by the advent of the X-ray FEL, Santra et al. [7] demonstrated theoretically that two-atomic site DCH states can indeed be probed by XTPPS. In their proof-of-principle simulation, they used organic *p*-aminophenol molecule and assumed that the molecule is irradiated by the 1.2-fs, 1 keV X-ray pulse containing $1.0 \times 10^{11}$ photons, focused to a width of 1 μm. For the proposed XTPPS, it is important to have a pulse significantly shorter than the core-hole lifetimes (e.g., 3 to 7 fs for F, O, N and C atoms). Then it is guaranteed that the second X-ray photon is absorbed by the single core-hole (SCH) state prepared by the first X-ray photon absorbed by the molecule. The sum of the kinetic energies of the first and second photoelectrons provides the double core-hole ionization potential (DIP) with two core holes. If the pulse width is longer than the core-hole lifetime, then Auger decay may occur prior to the absorption of the second photon and thus the DCH states may not be probed.

Recently, inspired by the work of Santra et al. [7], we theoretically investigated the DCH states of closed-shell molecules and proposed DCH electron spectroscopy for chemical analysis with the XTPPS experiment [19]. The DCH states of open-shell molecules are also possible and of interest. They may deserve a special attention since much more spin states are expected for the DCH states than those for closed-shell molecules and the trend of the relaxation energies can be different because of the valence open-shell electrons and many spin states in open-shell systems. However, no DCH states of open-shell molecules have ever been investigated so far. In this letter, we have thus examined DCH states, as well as SCH states, of open-shell molecules ($O_2$, NO, CN, and $NO_2$) using ab initio approach. Although there are more states in open-shell molecules than in closed-shell molecules, we will show that we can extract general trends common to both closed-shell and open-shell molecules. Using ab initio state energies, we show how to extract physically meaningful quantities, such as generalized intra- and interatomic relaxation energies, from the experimental spectra. We demonstrate that the DIPs of DCH states, which can be measured experimentally by XTPPS, are indeed more sensitive to chemical environment than the ionization potentials (IPs) of SCH states that can be probed by conventional X-ray photoelectron spectroscopy.

**2. Computational details**

Ab initio calculations of the vertical ionizations related to the SCH and DCH states of $O_2$, NO, CN, and $NO_2$ were performed. Molecular geometries were taken from experimental values [20,21]. For calculating the SCH and DCH states, complete active space self-consistent field (CASSCF) method (or ΔCASSCF) [22] was used and the configurations were restricted to those having one and two holes, respectively, in $K$-shell orbitals. The 1s orbitals were frozen using the Hartree-Fock (HF) MOs of neutral molecules and the active space was adopted as the full-valence orbitals which consist of all the possible bonds with 2s and 2p orbitals to account for the relaxations. The ΔSCF calculations which include 1~4 configurations depending on the character of the core-hole states were also performed to evaluate the relaxation energies. The restricted method was adopted in the SCF calculations. The correlation-consistent polarized valence triple zeta (cc-pVTZ) basis sets of Dunning [23] were adopted in the CASSCF and ΔSCF calculations of the core-hole states. We note that Koopmans' theorem in an open-shell system is not straightforward, as discussed in Refs. [24,25]. Plakhutin and Davidson et al. showed that the restricted open-shell Hartree-Fock (ROHF) method using the unified coupling operators, so-called as the canonical ROHF method, provides the orbital energies with which Koopmans' theorem holds in open-shell system [24,25]. Therefore, the ground-state orbital energies of $O_2$ and CN were calculated by the canonical ROHF method. For NO and $NO_2$, however, we could not obtain suitable solutions by canonical ROHF and therefore, the orbital energies were calculated by the MC(multi-configurational)-SCF method including the orbital localization and symmetry adaptation.

For $O_2$ and $NO_2$, we calculated the SCH states in both delocalized and localized molecular orbital pictures, but only the results based on the delocalized picture are presented in the next section. The interrelation between the localized and delocalized single core holes in molecules was studied before [26,27]. As for the DCH states of $O_2$ and $NO_2$, we calculated them in both delocalized and localized pictures following the recipe given by Cederbaum et al. [1]. In the CASSCF calculations, only the 1s orbitals of $O_2$ ($NO_2$) were localized by the Boys localization [28] restricting $C_{2v}$ ($C_s$) symmetry, and other orbitals were not localized. The molecular axis of $O_2$ was taken as the $C_2$ axis in $C_{2v}$ and the molecular plane was taken as the $C_s$ plane for $NO_2$. In the delocalized picture $D_{\infty h}$ ($C_{2v}$), the DCH states were described by $S_1^{-2} \pm S_2^{-2}$ and $S_1^{-1} S_2^{-1}$ where $S_1$ and $S_2$ are localized orbitals. For details; see Eqs. (4a)-(4d) in Ref. [1]. The localized DCH states $S_1^{-2}$ ($C_{\infty v}$) were calculated with the configurations that have double core holes at a one-atomic site.

The CASSCF and ΔSCF calculations were performed using the Molpro2008 quantum chemistry package [29]. The canonical ROHF calculations were carried out with the GAMESS program [30].

**3. Results and discussion**

*3.1. Single core-hole states*

In Table 1, we show the CASSCF and ΔSCF results for IPs with the experimental values [31]. The CASSCF values show reasonable agreement with the experimental data. The SCH states in the closed-shell molecules are always doublet states. In contrast, two different spin states always appear for the SCH states in open-shell molecules. For example, doublet and quartet states appear in the SCH states of $O_2$ and singlet and triplet states appear in the SCH states of NO, CN, and $NO_2$. The energy splitting of the spin states $\Delta E_{D-Q}$, i.e., the exchange energy, in $O_2$ calculated by CASSCF is 0.67 eV. This value is compared with the singlet-triplet splitting $\Delta E_{S-T}$ of other hetero-nuclear molecules NO and CN, 1.37 and 1.68 eV for N $1s^{-1}$ in NO and C $1s^{-1}$ in CN, respectively, while the splittings $\Delta E_{S-T}$ are 0.49 and 0.16 eV for O $1s^{-1}$ in NO and N $1s^{-1}$ in CN, respectively. These comparisons illustrate general tendency that the exchange interaction is larger (smaller) for the SCH state with the core hole located at the electron donating (accepting) atom in the heteronuclear molecule than those of the equivalent atom in the homonuclear molecule. It is also notable that the g-u splitting, i.e., the energy splitting between gerade and ungerade states, of the SCH states in $O_2$ is 42 meV for the quartet state, which is much larger than the splitting of 6 meV for the doublet state. These two values reasonably agree with the values calculated by the RAS (restricted active space)-SCF method, i.e., 50 meV for the quartet state and 7 meV for the doublet state [32]. Singlet-triplet splittings of $NO_2$ are 0.50 eV for O $1s^{-1}$ and 0.67 eV for N $1s^{-1}$, which are similar to $\Delta E_{D-Q}$ =0.67 eV for O $1s^{-1}$ of $O_2$. The g-u splittings in the O $1s^{-1}$ SCH states of $NO_2$ are much smaller than those of $O_2$ due to larger separation of the two O atoms in $NO_2$.

In Table 1, we also give the generalized relaxation energy $RC(S_i^{-1})$ which is approximated as

$$IP(S_i^{-1}) = -\varepsilon_{S_i} - RC(S_i^{-1}),  \qquad (1)$$

where $\varepsilon_{S_i}$ is an orbital energy of a neutral molecule. The orbital energies were calculated by the canonical ROHF method [24,25] for $O_2$ and CN, and by the MCSCF method for NO and $NO_2$. The orbital energies obtained by these two methods are also discussed in Appendix 1. The $RC(S_i^{-1})$ for O $1s^{-1}$ SCH states of heteroatomic molecules are ~19 eV, independent of molecule since the charge of O atom is scarcely different among these molecules. In contrast, $RC(S_i^{-1})$ for N $1s^{-1}$ SCH states vary depending on the chemical environment of the N atom. $RC(S_i^{-1})$ is larger when the N atom is next to the electron accepting O atom (in NO) than to the electron donating C atom (in CN) and increases with the number of the neighboring electron accepting O atom (from NO to $NO_2$). This is because the electron accepting O atoms can back-donate more electron densities to the N atom with the core hole than the electron donating C atom. These trends are common also for closed-shell molecules [19].

The generalized relaxation energy $RC(S_i^{-1})$ includes both orbital relaxation $R(S_i^{-1})$ and electron correlations $C(S_i^{-1})$

$$RC(S_i^{-1}) = R(S_i^{-1}) + C(S_i^{-1}) \qquad (2)$$

Generally, $R(S_i^{-1})$ is a dominant term and $C(S_i^{-1})$ is a minor correction in the core-electronic processes. We can approximately estimate these two contributions in the following manner: The CASSCF method takes account of the static correlations and the orbital reorganization due to core ionization, though it does not take the dynamic correlations into account. The ΔSCF, on the other hand, includes only the orbital relaxation effect. Thus, the differences between the results by CASSCF and those by ΔSCF may be approximated to the static correlations. The values of the correlation energy $C(S_i^{-1})$ estimated as the difference between CASSCF and ΔSCF are also given in Table 1.

The values of $C(S_i^{-1})$ for $O_2$ (~10 eV) are significantly larger than those for NO and CN (< 2 eV). Similar discrepancies also appear for the O $1s^{-1}$ SCH states in $NO_2$. These discrepancies can be ascribed to the failure of the ΔSCF method in the delocalized picture to account for all relaxation contributions [27]. This can be qualitatively understood in the point charge model that a delocalized hole has one-half of the screening energy of a localized hole [27]. To extract relaxation effects separately from the electron correlations for these cases, one has to adopt the localized picture in the ΔSCF calculations [1,19].

*3.2. Double core-hole states*

In Table 2, the results of DIPs are presented with the main configurations of the DCH states. Energy differences between the CASSCF and ΔSCF results, or electron correlation energies $C$, are also given. The correlation energies $C$ for the DCH states are in general larger than those for the SCH states, exhibiting that the influences of electron correlations on DIP are larger than those on IP.

Let us focus on the oxygen case as an example. In the localized representation, we obtained the DIPs of the one-site DCH states $O_1 1s^{-2}$ and $O_2 1s^{-2}$. Carrying out calculations with wave functions described by $O_1 1s^{-2} \pm O_2 1s^{-2}$ gives rise to DIPs in the delocalized picture. The extraordinary large electron correlation energy $C$ for $O_1 1s^{-2} \pm O_2 1s^{-2}$ is due to the failure of the ΔSCF method in the delocalized picture to account for all relaxation contributions [1,19] as the calculations for the SCH states [26,27] described in the previous subsection. We see a similar large discrepancy also for the $O_1 1s^{-2} \pm O_2 1s^{-2}$ in $NO_2$ due for the same reason. The energy difference between the $O1s^{-2}$ (1177eV) and $O_1 1s^{-2} \pm O_2 1s^{-2}$ (1182eV) states in the CASSCF results can be regarded as the deficiency of the relaxation in the delocalized picture. We thus refer the DIP values obtained in the localized picture in the rest of the paper.

The two-site DCH states $O_1 1s^{-1} O_2 1s^{-1}$ in $O_2$ are located at ~70 eV below the one-site states $O_1 1s^{-2}$ and $O_2 1s^{-2}$. The $O_1 1s^{-1} O_2 1s^{-1}$ $T_1$ state with the anti-parallel spins in the core orbitals is more stable than the $O_1 1s^{-1} O_2 1s^{-1}$ $T_2$ state with the anti-parallel spins in the valence π* orbitals. The $O_1 1s^{-1} O_2 1s^{-1}$ $S_1$ state has anti-parallel spins in both core and valence orbitals, while in the $O_1 1s^{-1} O_2 1s^{-1}$ $S_2$ state, the spins in the core and valence π* orbitals are both parallel. The energy differences of the DCH states among the same spin symmetry are about ~ 1.08 ($T_1$-$T_2$) and ~ 1.25 eV ($S_1$-$S_2$) for triplet and singlet states, respectively, while the singlet-triplet energy splitting is ~ 0.60 eV. The quintet state is the most stable in the two-site DCH states and is located 1.08 eV below the $T_1$ state.

In NO and CN, two doublet states and one quartet state exist for the two-site DCH states. In the $D_1$ state, the spins in the core orbitals are anti-parallel, while in the $D_2$ state, the spins in the core orbitals are parallel. The energy differences between these $D_1$ and $D_2$ states are relatively large, 2.53 and 3.74 eV for $N1s^{-1}O1s^{-1}$ of NO and $C1s^{-1}N1s^{-1}$ of CN, respectively. A quartet state is located below the doublet states.

We now explain the electron energy spectra to be observed in the XTPPS using NO as an example. Table 3 gives expected kinetic energies of photoelectrons with the photon energy of 1 keV. At energy of ~ 16 eV below the N 1s and O 1s single photon ionization main lines, we can observe the peaks corresponding to the second-step photoionization of the O1s$^{-1}$ and N1s$^{-1}$ single core-hole states, respectively, leading to the N1s$^{-1}$O1s$^{-1}$ two-site DCH states. Depending on the final-state spin configuration, only specific intermediate states are possible. For example, the O 1s$^{-1}$ S state leads to the N1s$^{-1}$O1s$^{-1}$ D$_1$ state, while the O 1s$^{-1}$ T state leads to the N1s$^{-1}$O1s$^{-1}$ D$_2$ and Q states.  At energies of about ~ 80 and ~ 90 eV below the N 1s and O 1s main lines, respectively, we can observe the peaks corresponding to the second-step photoionization leading to N 1s$^{-2}$ and O 1s$^{-2}$ one-site DCH states, respectively.  Thus, in total 14 lines appear in the open-shell molecule NO, in contrast to the fact that only 8 lines appear in the closed-shell molecule CO. Here, photoelectron satellite lines and Auger lines have been ignored.  In order to design the experiment, one has to adjust the photon energies to minimize the overlap of the listed photoelectron main lines with photoelectron satellites and Auger lines. In a preliminary simulation, we found that the intensity of XTPPS can be controlled by changing the photon energy. The energy levels of the SCH and DCH states are depicted in Fig. 1 together with the kinetic energies of photoelectrons to be observed.

The energy relations between the electron kinetic energies that can be measured experimentally, and the single and double core-hole ionization potentials, $IP(S_i^{-1})$ and $DIP(S_i^{-1}, S_j^{-1})$ that can be calculated by ab initio methods are summarized as follows:

$$\Delta E = KE(S_i^{-1}) - KE(S_i^{-1}, S_j^{-1}) = DIP(S_i^{-1}, S_j^{-1}) - IP(S_i^{-1}) - IP(S_j^{-1}), \qquad (3)$$

where $KE(S_i^{-1})$ and $KE(S_i^{-1}, S_j^{-1})$ are the electron kinetic energies ejected via the first-step ionization leading to the $S_i$ SCH state and via the second-step ionization from the $S_j$ SCH state to the $S_i S_j$ DCH state. $S_i$ and $S_j$ can be at the same atomic site ($i=j$) or at two different atomic sites ($i \ne j$). In the rest of this paper, we focus on the quantity $\Delta E$ that can be measured via the experiments.  For calculation of $\Delta E$, we use the lowest spin states for both the final and intermediate states.  Other two-photon ionizations with slightly different $\Delta E$ values are of course possible, but the discussion below will be unchanged even if these small energy differences are taken into account.

*3.3. One-site double core-hole states*

In Table 4, we present the values of $\Delta E1(S_i^{-2}) = DIP(S_i^{-1}, S_i^{-1}) - [IP(S_i^{-1}) + IP(S_i^{-1})]$ for the one-site DCH states for the open-shell molecules studied here. From $\Delta E1(S_i^{-2})$, we can extract the generalized relaxation energy.  The relation between $\Delta E1(S_i^{-2})$ and $RC(S_i^{-1})$ holds in the same manner as those for the closed-shell molecules and is derived by the second-order perturbation theory [19]. The energy $DIP(S_i^{-1}, S_i^{-1})$ needed to create two core holes at the same site $S_i$ is approximated to

$$DIP(S_i^{-1}, S_i^{-1}) = -2\varepsilon_{S_i} - RC(S_i^{-1}, S_i^{-1}) + V_{S_i S_i S_i S_i}. \qquad (4)$$

where $V_{S_i S_i S_i S_i}$ is one-site two-hole electron repulsion integral [1], that is $(ii|ii)$. Although additional terms also exist due to the open-shell structure, we approximate the relation between DIP and relaxation energy as Eq. (4). Here the DCH relaxation energy $RC(S_i^{-1}, S_i^{-1})$ is dominated by the relaxation energy and can be approximated to [1,2,19],

$$RC(S_i^{-1}, S_i^{-1}) = 4RC(S_i^{-1}). \qquad (5)$$

From Eqs. (1)–(5), we obtain the expression for the generalized relaxation energy,

$$RC(S_i^{-1}) = [\Delta E1(S_i^{-2}) - V_{S_i S_i S_i S_i}]/2. \qquad (6)$$

Single core-hole generalized relaxation energies $RC(S_i^{-1})$ were calculated based on Eq. (6). The CASSCF values of $RC(S_i^{-1}) = [\Delta E1(S_i^{-2}) - V_{S_i S_i S_i S_i}]/2$ are given in Table 4. For calculating these values, we temporarily used the lowest quartet state for $O_2$ and the lowest triplet states for NO, CN and $NO_2$. In principle, one can experimentally obtain these generalized relaxation energies from the measurable quantities $\Delta E1(S_i^{-2})$. The agreement between these two $RC(S_i^{-1})$ values, $RC(S_i^{-1})$ calculated from $[\Delta E1(S_i^{-2}) - V_{S_i S_i S_i S_i}]/2$ and $RC(S_i^{-1})$ estimated by Eq. (1) in Table 1, is reasonable with some deviation (<3 eV). This difference reflects the second-order result and also the effect of electron correlations as mentioned above [1,19]. This trend is similar to that in the closed-shell molecules [19].

*3.4. Two-site double core-hole states*

Let us consider the case of the two-site DCH states. Table 4 also includes the values of $\Delta E2(S_i^{-1}, S_j^{-1}) = DIP(S_i^{-1}, S_j^{-1}) - [IP(S_i^{-1}) + IP(S_j^{-1})]$ for the two-site DCH states. The $DIP(S_i^{-1}, S_j^{-1})$ of the two-site DCH states can be approximated to

$$DIP(S_i^{-1}, S_j^{-1}) = -\varepsilon_{S_i} - \varepsilon_{S_j} - RC(S_i^{-1}, S_j^{-1}) + V_{S_i S_i S_j S_j}. \qquad (7)$$

Here the double-vacancy relaxation energy $RC(S_i^{-1}, S_j^{-1})$ can be decomposed into three components:

$$RC(S_i^{-1}, S_j^{-1}) = RC(S_i^{-1}) + RC(S_j^{-1}) + IRC(S_i^{-1}, S_j^{-1}). \qquad (8)$$

Thus we have

$$IRC(S_i^{-1}, S_j^{-1}) = V_{S_i S_j S_i S_j} - \Delta E2(S_i^{-1}, S_j^{-1}). \qquad (9)$$

The explicit expression of the *interatomic* relaxation energy $IRC(S_i^{-1}, S_j^{-1})$ up to the second order in terms of the orbital energies and the two-electron integrals can be found in Refs. [1,2]. $V_{S_i S_j S_i S_j}$ is the two-site electron repulsion integral $(ii \mid jj)$ that can be well approximated to $1/R$, where $R$ is the nuclear distance. Table 4 also includes $IRC(S_i^{-1}, S_j^{-1})$ as calculated with Eq. (9). For calculating these values, we used the $T_1$ state for $O_2$, $D_1$ states for NO and CN, and $D_1$ and $D_2$ states for $NO_2$. In principle, one can experimentally obtain $IRC(S_i^{-1}, S_j^{-1})$ from the known or estimated distance $R$ between the two sites.

Let us now focus on diatomic molecules. We find that all of them have negative values for $IRC(S_i^{-1}, S_j^{-1})$, namely, –2.784, –3.444 and –1.999 eV, for $O_2$, NO and CN, respectively, as in the case of closed-shell molecules [19]. A negative value indicates suppression of the relaxation according to Eq. (8). The suppression of the relaxation for diatomic molecules seems to be a general rule. This suppression can be interpreted by considering the change in the electron density [19]. A core hole at $S_i$ attracts valence electrons and increases the electron density in the vicinity of $S_i$, yielding relaxation energy, and in return reduces the electron density in the vicinity of $S_j$. The relaxation energy for the second hole at $S_j$, having a core hole $S_i$, should therefore be reduced.

The $IRC(S_i^{-1}, S_j^{-1})$ for $NO_2$ exhibits an interesting behavior similar to the closed-shell triatomic molecules $CO_2$ and $N_2O$ [19]. The $IRC(N^{-1}, O^{-1})$ with two holes in adjacent atoms is positive, 0.788 eV. In this triatomic case, another O atom plays a role of an electron donor and enhances the relaxation of the DCH in two sites. The $IRC(O_1^{-1}, O_2^{-1})$ with two holes in the two oxygen terminal atoms is, on the other hand, is negative –2.113 eV. In this case, creation of the core hole on one site already withdraws the electron density from the central nitrogen atom and thus reduces the possibility of relaxation due to the creation of the second hole in the other terminal site. These trends seem to be valid for triatomic molecules, but this issue should be further examined carefully.

## 4. Summary

We have investigated the DCH states in open-shell molecules via CASSCF calculations, aiming to provide a guidance for designing an experiment for XTPPS.    Many spin states were obtained for the SCH and DCH states of open-shell molecules compared to closed-shell molecules and, therefore, more lines are expected in XTPPS for open-shell molecules.    This issue has been demonstrated in details for the XTPPS of NO. The trends of the generalized intra- and interatomic relaxation energies $RC(S_i^{-1})$ and $IRC(S_i^{-1}, S_j^{-1})$ of open-shell molecules are similar to those of closed-shell molecules. The $RC(S_i^{-1})$ and $IRC(S_i^{-1}, S_j^{-1})$ are related to the electron donating/withdrawing character of adjacent atoms and the connected chemical bonds.

**Acknowledgements**

The authors acknowledge Prof. L.S. Cederbaum for the valuable discussions and directions to the DCH electron spectroscopy. The authors also thank the reviewers for their helpful comments. M.E. acknowledges the support from JST-CREST and a Grant-in-Aid for Scientific Research from the Japan Society for the Promotion of Science, the Next Generation Supercomputing Project. K.U. acknowledges the support for the X-ray Free Electron Laser Utilization Research Project of Ministry of Education, Culture, Sports, Science and Technology of Japan (MEXT). The computations were partly performed using Research Center for Computational Science, Okazaki, Japan.

**Appendix 1**

The canonical ROHF [24,25] and MCSCF calculations have been performed for $O_2$ and CN to compare the ground-state orbital energies: the results of these two methods are compared, since no suitable canonical ROHF solutions could be obtained for NO and $NO_2$. The orbital energies calculated by the MCSCF method are –425.284(N1s) and –308.945(C1s) eV for CN, and –564.104 eV for O1s of $O_2$. The deviations of these orbital energies from those in Table 1 (canonical ROHF) are 0.06 and 0.38 eV for CN and 0.56 eV for $O_2$.    We believe that these differences do not affect the qualitative analysis made in the present work.

Table 1

Single core-hole IPs, generalized relaxation energies $RC(S_i^{-1})$ and correlation energies $C(S_i^{-1})$ calculated by the ΔSCF and CASSCF methods (in eV).

|  | Character [a] | $-\varepsilon_{S_i}$ | ΔSCF | CASSCF | | | Exptl.[b] |
|---|---|---|---|---|---|---|---|
|  |  |  | IP | IP | $RC(S_i^{-1})$ | $C(S_i^{-1})$ | IP |
| $O_2$ | $O1s^{-1}$, D, u | 563.545[c] | 555.340 | 545.574 | 17.971 | 9.766 | 544.47 |
|  | $O1s^{-1}$, D, g |  | 555.354 | 545.580 | 17.965 | 9.774 |  |
|  | $O1s^{-1}$, Q, u |  | 554.606 | 544.864 | 18.681 | 9.742 | 543.55 |
|  | $O1s^{-1}$, Q, g |  | 554.642 | 544.906 | 18.639 | 9.736 |  |
| NO | $O1s^{-1}$, S | 563.112[d] | 543.854 | 544.125 | 18.987 | −0.271 | 543.6 |
|  | $O1s^{-1}$, T |  | 543.362 | 543.634 | 19.478 | −0.272 | 543.1 |
|  | $N1s^{-1}$, S | 427.819[d] | 413.777 | 411.895 | 15.924 | 1.882 | 411.5 |
|  | $N1s^{-1}$, T |  | 412.396 | 410.525 | 17.294 | 1.871 | 410.1 |
| CN | $N1s^{-1}$, S | 425.222[c] | 408.894 | 407.851 | 17.371 | 1.043 | – |
|  | $N1s^{-1}$, T |  | 408.702 | 408.008 | 17.214 | 0.694 | – |
|  | $C1s^{-1}$, S | 308.561[c] | 298.325 | 296.328 | 12.233 | 1.997 | – |
|  | $C1s^{-1}$, T |  | 296.363 | 294.647 | 13.914 | 1.716 | – |
| $NO_2$ | $O1s^{-1}$, S, g | 562.682[d] | 554.284 | 543.622 | 19.060 | 10.662 | 542.0 |
|  | $O1s^{-1}$, S, u |  | 554.285 | 543.623 | 19.059 | 10.662 |  |
|  | $O1s^{-1}$, T, g |  | 553.776 | 543.118 | 19.564 | 10.658 | 541.3 |
|  | $O1s^{-1}$, T, u |  | 553.779 | 543.121 | 19.561 | 10.658 |  |
|  | $N1s^{-1}$, S | 431.879[d] | 414.724 | 413.466 | 18.413 | 1.258 | 413.3 |
|  | $N1s^{-1}$, T |  | 414.500 | 412.791 | 19.088 | 1.709 | 412.6 |

[a] S, D, T, and Q mean singlet, doublet, triplet, and quartet states, respectively. g and u denote gerade and ungerade states, respectively.
[b] Experimental values were taken from Ref. [31].
[c] Orbital energies were calculated by the canonical ROHF method [24,25]. For $O_2$, the averaged value of 563.561(g) and 563.528(u) is given.
[d] Orbital energies were calculated by the MCSCF method.

Table 2
Double core-hole IPs (DIPs) calculated by the ΔSCF and CASSCF methods (in eV).

| | State [a] | ΔSCF DIP | CASSCF DIP | $C$ [c] | Main configuration [b] |
|---|---|---|---|---|---|
| $O_2$ | $O_11s^{-2}$, T (L) | 1175.967 | 1177.289 | −1.322 | $(O_11s)^0$ |
| | $O_21s^{-2}$, T (L) | 1175.967 | 1177.289 | −1.322 | $(O_21s)^0$ |
| | $O_11s^{-2} - O_21s^{-2}$, T | 1223.415 | 1181.972 | 41.443 | $(O_11s)^0 - (O_21s)^0$ |
| | $O_11s^{-2} + O_21s^{-2}$, T | 1223.415 | 1181.972 | 41.443 | $(O_11s)^0 + (O_21s)^0$ |
| | $O_11s^{-1}O_21s^{-1}$, $T_1$ | 1106.776 | 1104.343 | 2.433 | $(O_11s)^\alpha(O_21s)^\beta(\pi_x^*)^\alpha(\pi_y^*)^\alpha$ |
| | $O_11s^{-1}O_21s^{-1}$, $T_2$ | | 1105.418 | 1.358 | $(O_11s)^\alpha(O_21s)^\alpha(\pi_x^*)^\alpha(\pi_y^*)^\beta$ |
| | $O_11s^{-1}O_21s^{-1}$, $S_1$ | 1105.117 | 1104.945 | 0.172 | $(O_11s)^\alpha(O_21s)^\beta(\pi_x^*)^\alpha(\pi_y^*)^\beta$ |
| | $O_11s^{-1}O_21s^{-1}$, $S_2$ | | 1106.196 | −1.079 | $(O_11s)^\alpha(O_21s)^\alpha(\pi_x^*)^\beta(\pi_y^*)^\beta$ |
| | $O_11s^{-1}O_21s^{-1}$, Quintet | 1104.456 | 1103.267 | 1.189 | $(O_11s)^\alpha(O_21s)^\alpha(\pi_x^*)^\alpha(\pi_y^*)^\alpha$ |
| NO | $O1s^{-2}$, D | 1176.435 | 1177.704 | −1.269 | $(O1s)^0$ |
| | $N1s^{-2}$, D | 904.855 | 902.947 | 1.908 | $(N1s)^0$ |
| | $N1s^{-1}O1s^{-1}$, $D_1$ | 973.159 | 970.114 | 3.045 | $(O1s)^\alpha(N1s)^\beta(\pi^*)^\alpha$ |
| | $N1s^{-1}O1s^{-1}$, $D_2$ | | 972.645 | 0.514 | $(O1s)^\alpha(N1s)^\alpha(\pi^*)^\beta$ |
| | $N1s^{-1}O1s^{-1}$, Q | 972.550 | 970.019 | 2.531 | $(O1s)^\alpha(N1s)^\alpha(\pi^*)^\alpha$ |
| CN | $N1s^{-2}$, D | 894.884 | 895.339 | −0.455 | $(N1s)^0$ |
| | $C1s^{-2}$, D | 663.627 | 661.649 | 1.978 | $(C1s)^0$ |
| | $C1s^{-1}N1s^{-1}$, $D_1$ | 721.285 | 716.893 | 4.392 | $(N1s)^\alpha(C1s)^\beta(\sigma)^\alpha$ |
| | $C1s^{-1}N1s^{-1}$, $D_2$ | | 720.632 | 0.653 | $(N1s)^\alpha(C1s)^\alpha(\sigma)^\beta$ |
| | $C1s^{-1}N1s^{-1}$, Q | 721.177 | 717.481 | 3.696 | $(N1s)^\alpha(C1s)^\alpha(\sigma)^\alpha$ |
| $NO_2$ | $O_11s^{-2}$, D (L) | 1172.206 | 1171.713 | 0.493 | $(O_11s)^0$ |
| | $O_21s^{-2}$, D (L) | 1172.206 | 1171.713 | 0.493 | $(O_21s)^0$ |
| | $O_11s^{-2} - O_21s^{-2}$, D | 1219.209 | 1177.139 | 42.070 | $(O_11s)^0 - (O_21s)^0$ |
| | $O_11s^{-2} + O_21s^{-2}$, D | 1219.209 | 1177.139 | 42.070 | $(O_11s)^0 + (O_21s)^0$ |
| | $O_11s^{-1}O_21s^{-1}$, $D_1$ | 1096.804 | 1095.914 | 0.890 | $(O_11s)^\alpha(O_21s)^\beta(\pi^*)^\alpha$ |
| | $O_11s^{-1}O_21s^{-1}$, $D_2$ | 1097.217 | 1096.602 | 0.615 | $(O_11s)^\alpha(O_21s)^\alpha(\pi^*)^\beta$ |
| | $O_11s^{-1}O_21s^{-1}$, Q | 1096.527 | 1095.73 | 0.797 | $(O_11s)^\alpha(O_21s)^\alpha(\pi^*)^\alpha$ |
| | $N1s^{-2}$, D | 905.366 | 904.154 | 1.212 | $(N1s)^0$ |
| | $N1s^{-1}O1s,g^{-1}$, $D_1$ | 968.874 | 968.033 | 0.841 | $(O_11s, g)^\alpha(N1s)^\beta(\pi^*)^\alpha$ |
| | $N1s^{-1}O1s,u^{-1}$, $D_1$ | | 968.036 | 0.838 | $(O_11s, u)^\alpha(N1s)^\beta(\pi^*)^\alpha$ |
| | $N1s^{-1}O1s,g^{-1}$, $D_2$ | 969.085 | 968.367 | 0.718 | $(O_11s, g)^\alpha(N1s)^\alpha(\pi^*)^\beta$ |
| | $N1s^{-1}O1s,u^{-1}$, $D_2$ | | 968.367 | 0.718 | $(O_11s, u)^\alpha(N1s)^\alpha(\pi^*)^\beta$ |
| | $N1s^{-1}O1s,g^{-1}$, Q | 981.394 | 967.855 | 13.539 | $(O_11s, g)^\alpha(N1s)^\alpha(\pi^*)^\alpha$ |
| | $N1s^{-1}O1s,u^{-1}$, Q | 981.542 | 967.858 | 13.684 | $(O_11s, u)^\alpha(N1s)^\alpha(\pi^*)^\alpha$ |

[a] S, D, T, and Q mean singlet, doublet, triplet, and quartet states, respectively. (L) means solution with localized picture.
[b] Main configurations are shown with only open-shell orbitals.
[c] $C$ denotes the difference between ΔSCF and CASSCF values.

Table 3

Kinetic energies (KE) of photoelectrons in the XTPPS experiment for NO molecule by an X-ray pulse with photon energies of 1 keV.

| State [a] | KE (eV) | Intermediate State |
|---|---|---|
| $N1s^{-1}$, S | 588.105 | |
| $N1s^{-1}$, T | 589.475 | |
| $N1s^{-2}$ | 508.948 | $N1s^{-1}$, S |
| | 507.578 | $N1s^{-1}$, T |
| $N1s^{-1}O1s^{-1}$, $D_1$ | 573.520 | $O1s^{-1}$, T |
| $N1s^{-1}O1s^{-1}$, $D_2$ | 571.480 | $O1s^{-1}$, S |
| $N1s^{-1}O1s^{-1}$, Q | 573.615 | $O1s^{-1}$, T |
| $O1s^{-1}$, S | 455.875 | |
| $O1s^{-1}$, T | 456.366 | |
| $O1s^{-2}$ | 366.421 | $O1s^{-1}$, S |
| | 365.930 | $O1s^{-1}$, T |
| $N1s^{-1}O1s^{-1}$, $D_1$ | 440.411 | $N1s^{-1}$, T |
| $N1s^{-1}O1s^{-1}$, $D_2$ | 439.250 | $N1s^{-1}$, S |
| $N1s^{-1}O1s^{-1}$, Q | 440.615 | $N1s^{-1}$, T |

[a] S, D, T, and Q refer to singlet, doublet, triplet, and quartet states, respectively.

Table 4

Calculated energy differences $\Delta E1(S_i^{-2})$ and $\Delta E2(S_i^{-1}, S_j^{-1})$ with intra- and interatomic generalized relaxation energies $RC(S_i^{-1})$ and $IRC(S_i^{-1}, S_j^{-1})$ (eV).

| Molecule | Energy difference | | Generalized relaxation energy | |
|---|---|---|---|---|
| $O_2$ | $\Delta E1\,(O1s^{-2})$ | 87.561 | $RC(O1s^{-1})$ | 18.663 |
| | $\Delta E2\,(O_11s^{-1}, O_21s^{-1})$ | 14.615 | $IRC(O_11s^{-1}, O_21s^{-1})$ | –2.784 |
| NO | $\Delta E1\,(N1s^{-2})$ | 81.897 | $RC(N1s^{-1})$ | 13.328 |
| | $\Delta E1\,(O1s^{-2})$ | 90.436 | $RC(O1s^{-1})$ | 17.225 |
| | $\Delta E2\,(N1s^{-1}, O1s^{-1})$ | 15.955 | $IRC(N1s^{-1}, O1s^{-1})$ | –3.444 |
| CN | $\Delta E1\,(C1s^{-2})$ | 72.355 | $RC(C1s^{-1})$ | 9.933 |
| | $\Delta E1\,(N1s^{-2})$ | 79.323 | $RC(N1s^{-1})$ | 14.615 |
| | $\Delta E2\,(C1s^{-1}, N1s^{-1})$ | 14.238 | $IRC(C1s^{-1}, N1s^{-1})$ | –1.949 |
| $NO_2$ | $\Delta E1\,(N1s^{-2})$ | 78.572 | $RC(N1s^{-1})$ | 14.991 |
| | $\Delta E1\,(O1s^{-2})$ | 85.477 | $RC(O1s^{-1})$ | 19.705 |
| | $\Delta E2\,(N1s^{-1}, O1s^{-1})$ | 11.279 | $IRC(N1s^{-1}, O1s^{-1})$ | 0.788 |
| | $\Delta E2\,(O_11s^{-1}, O_21s^{-1})$ | 8.670 | $IRC(O_11s^{-1}, O_21s^{-1})$ | –2.113 |

Fig. 1. Energy levels of the SCH and DCH states (solid horizontal lines) and the observed photoelectrons (arrows) in the XTPPS with the photon energy of 1 keV ($h\nu_1$ and $h\nu_2$) demonstrated for NO.